# High-Resolution Imaging in the Visible on Large Ground-Based Telescopes.

Craig Mackay*[a], Rafael Rebolo[,e], Jonathan Crass[a], David L. King[a], Lucas Labadie[d], Víctor González-Escélera[b], Marta Puga[b], Antonio Pérez-Garrido[c], Roberto López[b], Alejandro Oscoz[b], Jorge A. Pérez-Prieto[b], Luis F. Rodríguez-Ramos[b], Sergio Velasco[b], Isidro Villó[c],

[a]Institute of Astronomy, University of Cambridge, Madingley Road, Cambridge CB3 0HA, UK
[b]Instituto de Astrofisica de Canarias, C/ Vía Láctea s/n, La Laguna, Tenerife E-38205, Spain and Departamento de Astrofísica, Universidad de La Laguna, La Laguna, Spain
[c]Universidad Politecnica de Cartagena, Campus Muralla del Mar, Cartagena, Murcia E-30202, Spain
[d]I. Physikalsiches Institut, Universität zu Köln, Zülpicher Strasse 77, 50937 Köln, Germany 38700
[e]Consejo Superior de Investigaciones Científicas, Spain

**ABSTRACT**

Lucky Imaging combined with a low order adaptive optics system has given the highest resolution images ever taken in the visible or near infrared of faint astronomical objects. This paper describes a new instrument that has already been deployed on the WHT 4.2m telescope on La Palma, with particular emphasis on the optical design and the predicted system performance. A new design of low order wavefront sensor using photon counting CCD detectors and multi-plane curvature wavefront sensor will allow virtually full sky coverage with faint natural guide stars. With a 2 x 2 array of 1024 x 1024 photon counting EMCCDs, AOLI is the first of the new class of high sensitivity, near diffraction limited imaging systems giving higher resolution in the visible from the ground than hitherto been possible from space.

**Keywords:** Lucky Imaging, adaptive optics, Charge coupled devices, EMCCDs, low light level imaging, nlCWFS.

## 1. INTRODUCTION

The Hubble Space Telescope (HST) has revolutionised many branches of astronomy by delivering much higher angular resolution than is possible with conventional ground-based telescopes. Although large telescopes are capable of much better angular resolution it has been very difficult to achieve major improvements. Most progress has been made in the near infrared on large (4-10 m class) telescopes. In the visible only one technique has demonstrated Hubble resolution on a Hubble sized (2.5 m diameter) telescope on the ground. This technique is known as Lucky Imaging. It was suggested originally by Hufnagel[1] in 1966 and given its name by Fried[2] in 1978. Images are recorded at high frame rates to freeze the motion caused by atmospheric turbulence. A relatively bright reference star in the field allows image quality to be determined. The best fraction of images are shifted and added to give a combined image close to diffraction limited in their image quality when a fraction of ~5-30% is selected, the percentage depending on the atmospheric conditions at the time.

The Lucky Imaging technique has become viable in recent years principally because of the development of electron multiplying CCDs (EMCCDs) particularly by E2V Technologies Ltd (Chelmsford, UK). These devices have most of the characteristics of conventional CCDs used routinely by astronomers and which are read out at low speed. However, a modification of the device output register[3] provides internal amplification by large factors, up to a few thousand times. At high pixel rates of up to 30 MHz and readout noise of around 100 electrons RMS this gain allows individual photons to be detected with good signal-to-noise. This permits imaging at high frame rates without any read noise penalty. In photon counting mode it is possible to operate with almost all of the quantum efficiency of a conventional CCD making these devices particularly attractive for a range of high-speed imaging and spectroscopy applications in astronomy and other research areas.

*cdm <at> ast.cam.ac.uk

In principle, large telescopes can deliver sharper images than HST. However, the probability of the Lucky Imaging technique delivering near diffraction limited images becomes vanishingly small for telescopes significantly larger than the HST[2]. This is because the number of turbulent cells across the diameter of the telescope is too large for there to be a significant chance of a relatively flat wavefront (and hence a near diffraction-limited image) across the aperture of the telescope. By increasing the diameter of the telescope we bring in the effects of yet larger scales of atmospheric turbulence. In principle, if we could eliminate the largest turbulent scales where most of the power in the atmospheric turbulence resides[4] then the probability of recording a sharp image will increase. Essentially, eliminating one turbulent scale reduces the phase variance across that scale so that the characteristic cell size, $r_0$ (defined as the scale size over which the variance is ~1 radian$^2$) is increased. Provided enough of the large turbulent scales are removed, the corrected $r_0$ will be large enough so the number of cells across the diameter of the telescope is similar to those typically encountered with an uncorrected 2.4 m aperture, the largest diameter for which the Lucky Imaging technique normally gives good results.

In order to demonstrate this, one of the Cambridge Lucky Imaging cameras was mounted on the Palomar 5-m telescope behind the low order adaptive optics system PALMAO[5]. We obtained images with a resolution more than 3 times that of the HST. The performance of this combination of Lucky Imaging plus low order adaptive optics is compared with that of the Advanced Camera for Surveys on HST in Figure 1. Lucky Imaging has now been used by a number of groups both in Europe and in the US and this has contributed to the scientific development of optical Lucky Imaging[6,7,8,9]. For example, FastCam and AstraLux have worked on 2-m and 4-m class telescopes achieving full diffraction-limited resolution in the z'- and I- bands. In the US, projects aiming at developing visible AO systems are also continuing [10]. New post-processing approaches have also been successfully implemented to further improve on the image sharpness and quality of the delivered images, reaching Strehl ratio significantly higher than possible with classical LI[11,12].

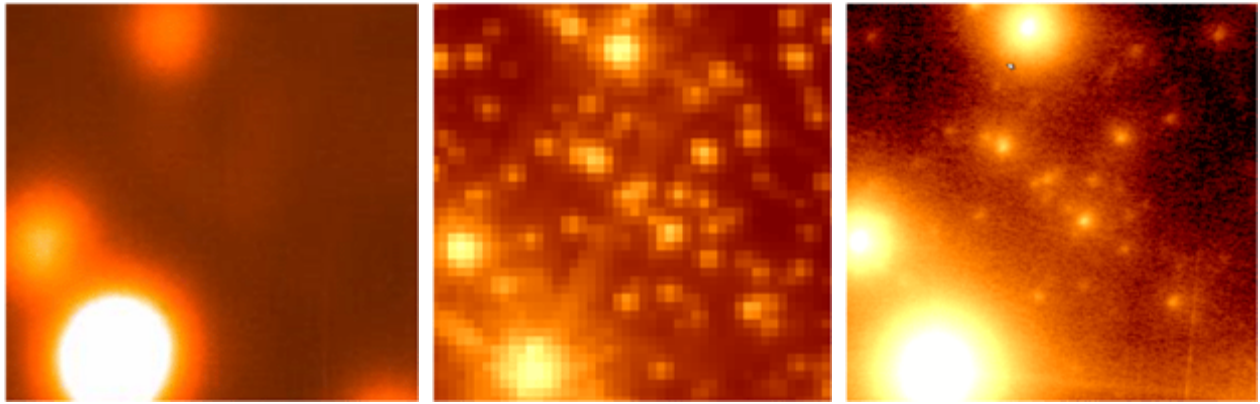

Figure 1: Comparison images of the core of the globular cluster M 13. On the left the image with natural (~0.65 arcsec) seeing on the Palomar 5-m telescope, the Hubble Advanced Camera for Surveys with ~120 milliarcsecond resolution (middle) and the Lucky Camera plus Low-Order AO image with 35 milliarcsecond resolution (right).

In common with most Shack-Hartmann based AO systems, PALMAO requires a very bright reference star for the Shack-Hartmann wavefront sensor. Shack-Hartmann AO systems generally need reference stars of I~12-14 magnitude, and these are very scarce[13]. AO systems that try to recover every moment on the telescope also have a very small isoplanatic patch in the visible of only a few arcseconds in diameter[14]. The net effect is that conventional Shack-Hartmann AO based systems can only be used over a very small fraction of the sky, < 0.1%. Laser guide stars have problems in delivering images with ~0.1 arcseconds accuracy. The consequence of these considerations is that it is very hard to achieve a resolution better than 0.1 arcseconds even in the near infrared on large telescopes although this has been achieved in a limited number of instances, particularly on very bright targets.

A study by Racine[15] showed that curvature sensors actually deployed on telescopes are significantly more sensitive than Shack-Hartmann sensors particularly when used for relatively low order turbulent correction. Olivier Guyon[16] simulated the performance of pupil plane curvature sensors and showed that they are very attractive in general and, when combined with EMCCDs, ought to give a substantial improvement in sensitivity over Shack-Hartmann sensors. This would allow operation over a much larger fraction of the sky. We have designed a new instrument, AOLI (Adaptive Optics assisted Lucky Imager), to let us carry out science combining Lucky Imaging and low order adaptive optics on targets over much of the sky. AOLI is a collaboration between the Instituto de Astrofisica de Canarias/Universidad de La Laguna

(Tenerife, Spain), the Universidad Politecnica de Cartagena (Spain), Universität zu Köln (Germany), the Isaac Newton Group of Telescopes (La Palma, Spain) and the Institute of Astronomy in the University of Cambridge (UK). This paper describes the current status of the instrument, its optical configuration and performance.

## 2.    AOLI: GENERAL CONFIGURATION

The AOLI instrument consists of a non-linear curvature wavefront sensor and a low order adaptive optics wavefront corrector using a deformable mirror in conjunction with a wide-field, array detector Lucky Imaging camera. It is designed specifically for use on the WHT 4.2-m and the GTC 10.4-m telescopes on La Palma but it could be used on almost any large telescope without major modification.

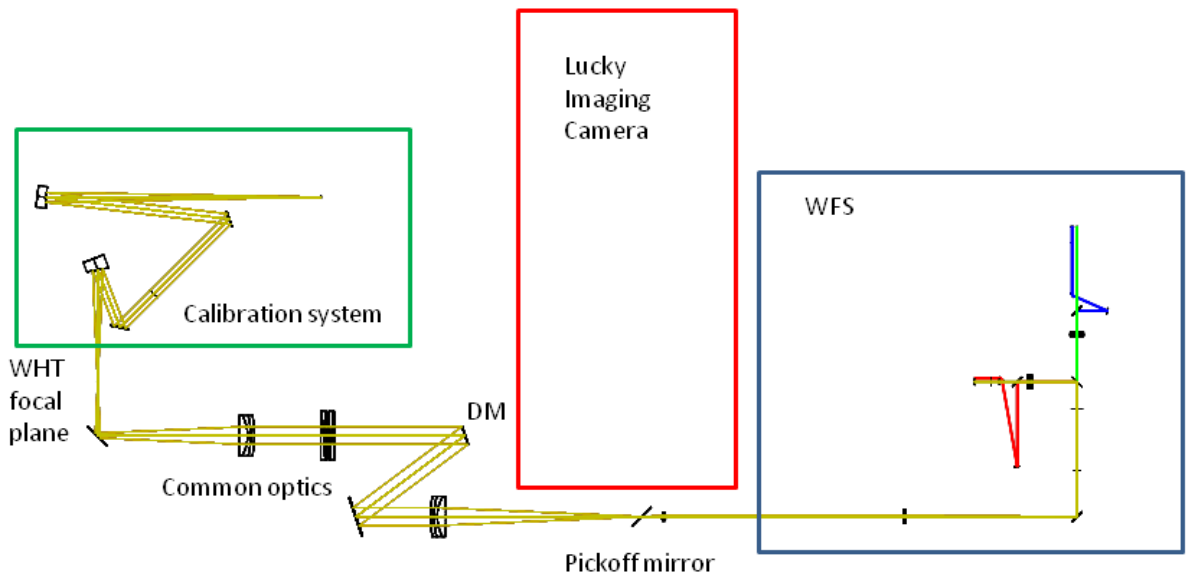

Figure 2: The optical layout of AOLI. Light enters AOLI from the bottom left-hand corner after passing through an image de-rotator mounted on the Naysmith instrument ring before reaching the telescope focus. The light is collimated and passed through an atmospheric dispersion corrector before it strikes a deformable mirror (DM). We selected the ALPAO-DM241 unit as it has excellent stability and provides an unusually long stroke. The light is then reflected and, after a fold mirror, is reimaged on to a pickoff mirror. The pickoff mirror deflects light towards the science camera (Figure 4). The light from the reference star goes directly on to the wavefront sensor (WFS). The telescope pupil is reimaged down to approximately 2 mm diameter from its original 4.2 m. Figure 3 shows the details of the wavefront sensor arrangement. In addition we have a calibration system which is designed to provide calibration sources that are of the same diameter in the telescope focal plane as the diffraction limit of the principal telescope. This is needed particularly to feed the wavefront sensor for setup and alignment. The wavefront sensor uses broadband light between 500 and 950 nm and therefore the calibration system has to be achromatic. Rather than using any refractive optics it uses off-axis paraboloid mirrors.

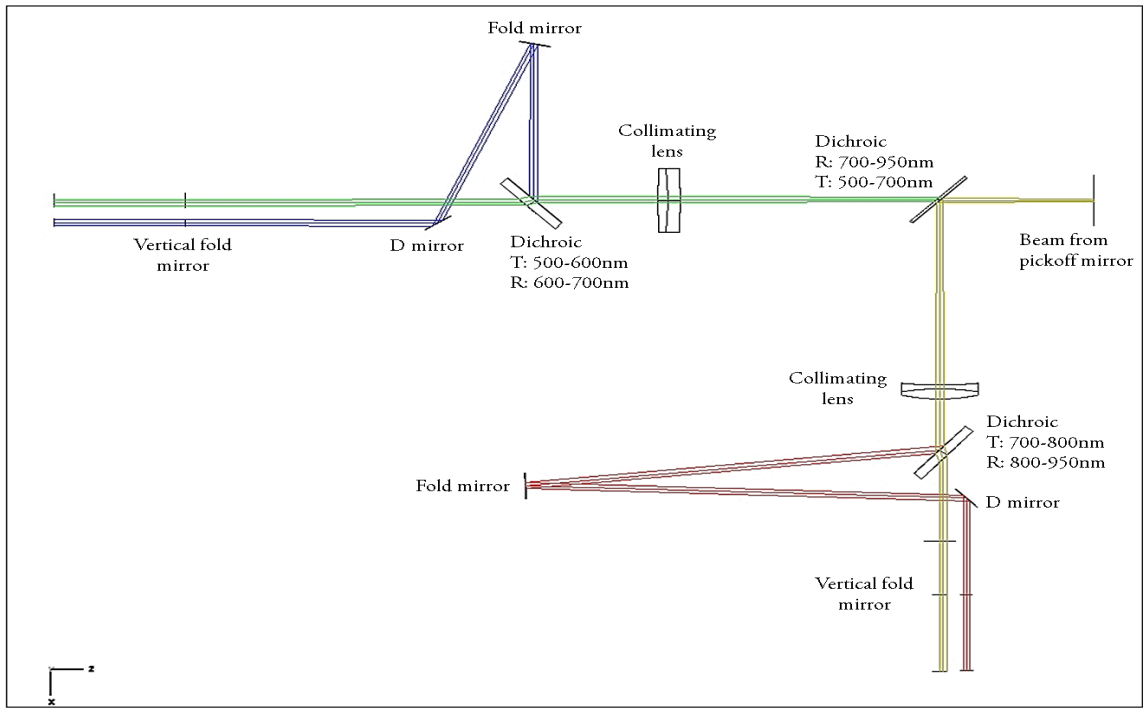

Figure 3: A model of the optical arrangement of the curvature sensor. The common optics take light from the WHT focus to the pickoff mirror where the light for the science camera is diverted to the science camera assembly. The light to the wavefront sensor passes through the pickoff mirror where is split into 2 beams with a dichroic beam splitter working at approximately 700 nm. Each beam is then split again with the second dichroic, one at 600 nm giving pupil images at wavelengths of ~550 nm and ~650 nm for one pair of beams and, with another dichroic beam splitter centred on about 800 nm gives a pair of pupil images at wavelengths of ~750 nm and ~870 nm. The pupil images are recorded with a pair of photon counting electron multiplying CCD cameras developed in Cambridge running at about 100 Hz frame rate. The data from these detectors are processed to derive the wavefront errors and these values are then used to drive the deformable mirror to remove the largest scales of turbulence from the wavefront. The science beam is turned 90° by the pickoff mirror plane to the science imaging detector (see Figure 4) which is then used in conventional Lucky Imaging mode.

AOLI includes an atmospheric dispersion corrector, essential for work away from the zenith, after the collimating lens and before the deformable mirror (the ADC is not marked in the figure). In Figure 2 the science beam goes from the telescope to a pickoff mirror which deflects light from the science beam towards the science camera shown in figure 4. The reference star is located on the optical axis of the telescope and passes through the pickoff mirror to the wavefront sensor. The light from the telescope is reflected via a deformable mirror set in a pupil plane of the telescope which allows the curvature errors determined by the curvature sensor sub-system to be corrected directly. The deformable mirror is manufactured by ALPAO (France) with 241 elements over the pupil. It will allow correction of wavefront errors on scales of > ~0.5m on the 4.2 m diameter WHT telescope. Our simulations[17] suggest that this will then give us a Lucky Imaging selection percentage under typical/good conditions of about 25-30% in I band. In addition, our simulations suggest that this deformable mirror will have a high enough resolution to achieve satisfactory correction on the GTC 10.4 m telescope. Our approach is to develop a system optimised for the WHT that may be modified and re-deployed quickly in order to demonstrate the technologies as convincingly as possible. Its subsequent deployment to the GTC will then follow.

The science camera (Figure 4) is a simple magnifier using custom optics to give diffraction limited performance. The camera is optimised for the 500nm to 1 micron wavelength range. The diffraction limit of the WHT (GTC) at 0.8μm (I band) is about 40 (15) mas and the camera offers a range of pixel scales of between 15 and 55 mas on the WHT. The camera uses an array of 4 photon counting, electron multiplying, back illuminated CCD201s manufactured by E2V Technologies Ltd, each 1024 x 1024 pixels. As the CCDs are non-buttable we use an arrangement similar to that of the original HST WF/PC (see Figure 4). Four small contiguous mirrors in the focal plane are slightly tilted and then individually reimaged on to a separate CCD. Each CCD has its own filter wheel. This allows the use of a narrowband filter, for example, for the science object with a broad filter for the reference star. The configuration allows a contiguous region of 2000 x 2000 pixels giving a field of view of from 112 x 112 arcsec down to 30 x 30 arcseconds on the WHT, depending on the magnification selected.

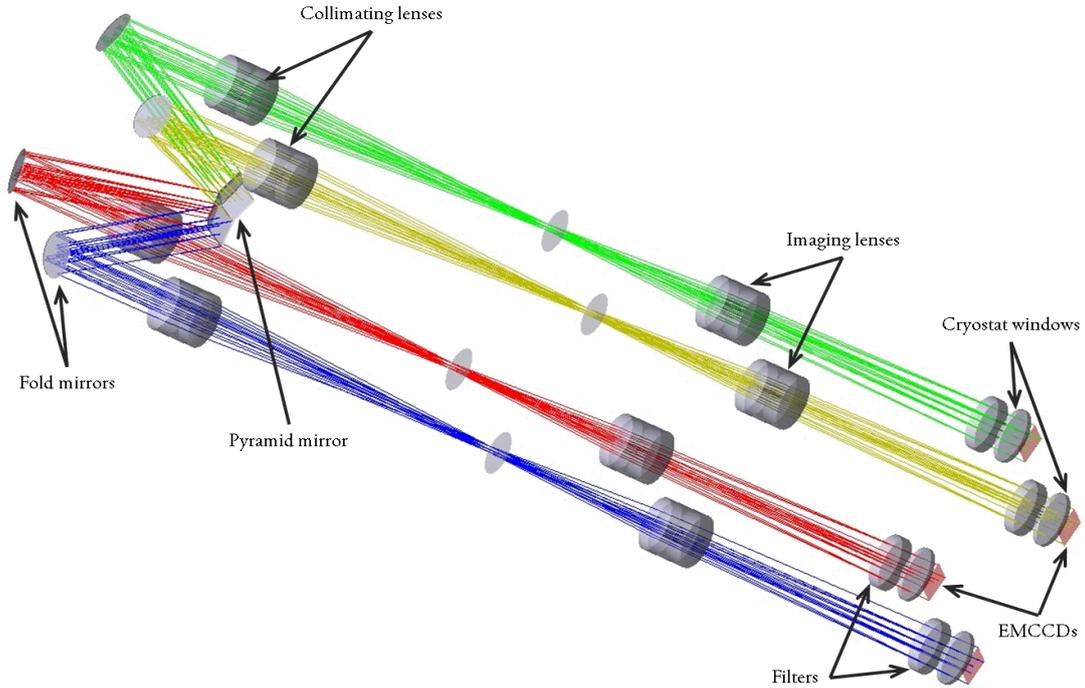

Figure 4: The science camera optical arrangement whereby the light from a single area of sky is split on to four separated and non-buttable CCDs. The magnified image (optics not shown) of the sky is projected onto a pyramid of four mirrors that reflects the light on to relay mirrors and via reimaging optics onto four electron multiplying detectors. This structure was suggested by the design of the original widefield/planetary camera installed on the HST.

The CCDs are back illuminated (thinned) with very high quantum efficiency (peak >95%) from E2V Technologies. Custom electronics developed in Cambridge give up to 30 MHz pixel rate, and 25 frames per sec. Higher frame rates are possible. A readout format of 2000 x 100 pixels gives ~200 fps, allowing high time resolution astronomy as well with the instrument. The data (~220 MBytes/sec continually) are streamed via the host computer to high-capacity RAID disk drive systems after lossless compression. The host computer performs basic lucky imaging selection, allowing image quality to be assessed while the exposure is progressing. The construction of the unit has been kept relatively low-cost. The instrument will be mounted at Nasmyth focus on an optical bench behind the WHT image rotator. On the GTC it is probable that the instrument would be mounted on one of the folded Cassegrain ports.

## 3. NON-LINEAR PHOTON COUNTING CURVATURE WAVEFRONT SENSORS

Adaptive optic systems most frequently depend on Shack-Hartmann wavefront sensors which require very bright and scarce reference stars (typically I~12-14 magnitude). Low order correction may be achieved in principle with much fainter reference stars. Curvature wavefront sensors work by taking images on either side of a conjugate pupil plane and by measuring changes in the intensity of illumination as the wavefront passes through the pupil. A part of the wavefront that becomes fainter as it goes through the pupil must correspond to a part of the wavefront that is diverging while if it becomes brighter it is converging. Racine[15] has shown that curvature sensors actually deployed on telescopes are typically 10 times (2.5 magnitudes) more sensitive than Shack-Hartmann sensors for the same degree of correction. In addition they are very much more sensitive again when used for low order correction as the system cell size is dynamically increased[8] and the wavefront sensor readout rate/integration time may be significantly reduced. Correction with coarse cell sizes allows averaging sensor signals over significant areas of the curvature sensor.

AOLI uses a more advanced version of this approach, the non-linear curvature wavefront sensor that uses four pupil planes to provide a more rapid and faster convergence in achieving a wavefront fit. The angular resolution we should achieve on the WHT/GTC will be typically 40/15 milliarcseconds in the visible to I-band range. Our simulations suggest that we should be able to operate at the faintest level of the reference star of about 17.5-18.0 on the WHT and about 18.5-19.0 mag on the GTC. This will allow us to find reference stars over nearly all the sky even at high galactic latitudes (>85%)[13]. At the fainter end of these ranges it is probable that we will only be able to achieve partial correction but enough to improve the variance of the wavefront entering the science camera. The AOLI wavefront sensor is described in more detail in another paper in this conference from Crass at al[18].

## 4. AOLI CALIBRATION SYSTEM

Any instrument design must take particular care to provide proper calibration facilities that allow the instrument to be set up fully in the laboratory or on the telescope during the daytime. The most demanding feature of AOLI is the need to generate an artificial star at the telescope focal plane no bigger than the diffraction limit of the telescope. This is further complicated because the curvature wavefront sensor design described above and by Crass et al.[18], uses light between 500 and 950 nm making the use of refractive optics unrealistically difficult. Because of this, we have developed an optical arrangement which uses a pair of off-axis paraboloids to give the necessary image quality. The outline design is shown in figure 5. The calibration system is described in more detail in another paper presented as part of this conference (Antolin et al [19]).

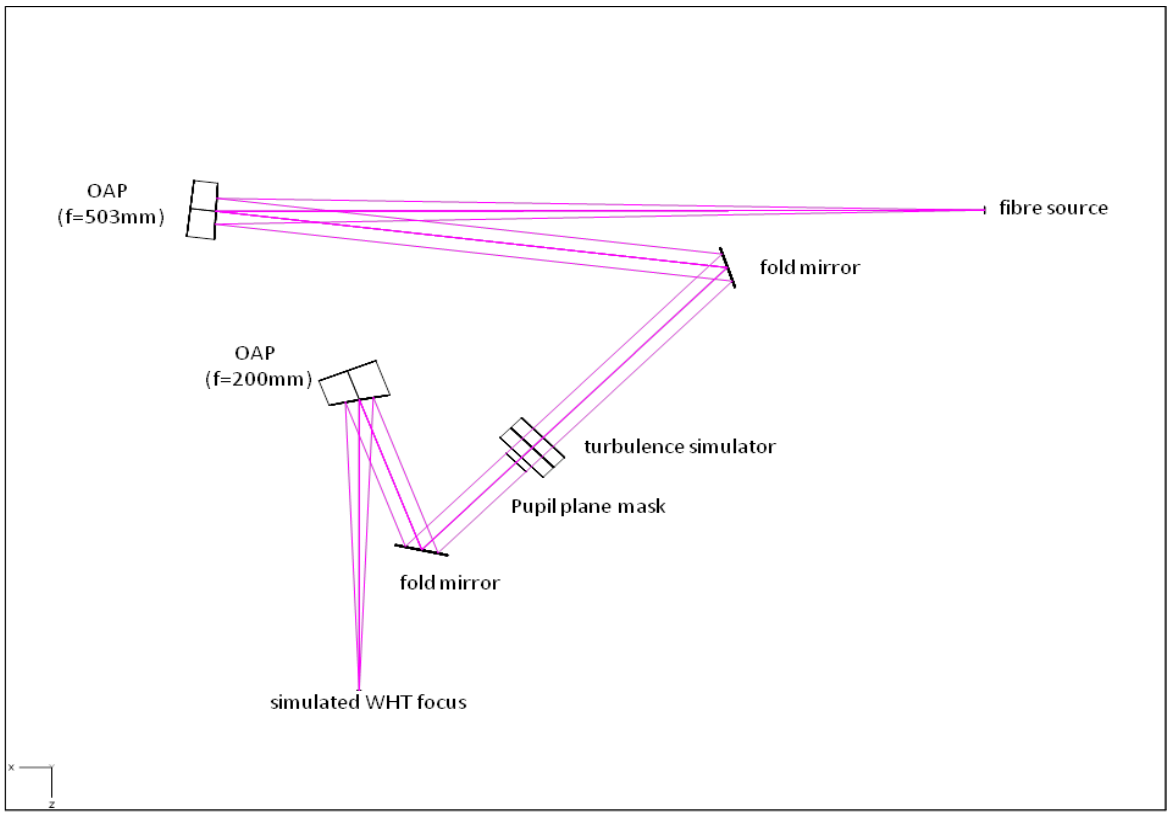

Figure 5: the AOLI calibration system uses a fibre source to generate a broadband input to a 2.5: 1 demagnification optical arrangement that uses a pair of off-axis paraboloids to produce a compact image at the same position as the focus of the telescope but can be used to feed the wavefront sensor and science camera optics of AOLI. It turbulence simulator is located close to the pupil plane of the reimaging system (OAP design from David King).

## 5. FIRST LIGHT EXPERIENCES WITH AOLI

The first observing run with AOLI was on the 4.2 m William Herschel telescope (WHT) in September 2013. In most respects the instrument worked well and Figure 6 shows it mounted at the Naysmith focus of the WHT. Unfortunately the weather conditions while at the telescope were remarkably poor with very poor seeing, high humidity and frequent dome closures. As a consequence very little in the way of scientific results were actually obtained. Nevertheless, a paper is in preparation describing results that were obtained ( Velasco et al.[20]).

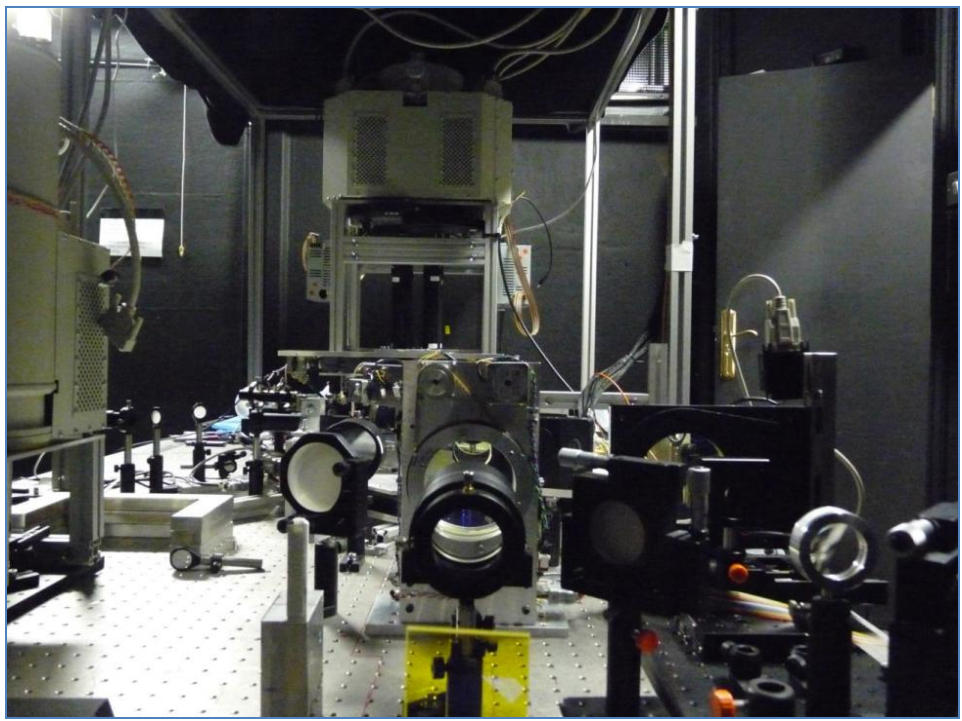

Figure 6: AOLI mounted at the Naysmith focus of the William Herschel 4.2 m Telescope in September 2013. Light comes from the telescope through the collimating optics immediately at the front of the picture. The deformable mirror is to the right and one of the curvature wavefront sensor EMCCD cameras is at the edge of the picture on the left. The science camera is vertically mounted at the far end of the instrument. Most of the light baffles used in the instrument were removed before this photograph was taken.

## 6. CONCLUSIONS

For many years instrument developers have been trying to produce an instrument capable of observing close to the diffraction limit of large telescopes in the visible from the ground, over the full sky with good observing efficiency. High-speed, high efficiency, photon counting detector availability has transformed our capacity to build instruments capable of working with the rapidly changing atmosphere and using very faint reference stars to let us optimise our recording of light from the sky. The combination of these EMCCDs with low order curvature wavefront sensors (also using EMCCDs) will allow a new generation of astronomers to explore the Universe with as big a step change in resolution as Hubble provided over 20 years ago. Hubble provided an eight-fold improvement over the typical ground-based image resolution of ~1 arcsec to give images of ~0.12 arcsec resolution. We have already demonstrated a further improvement over Hubble with the Palomar 5 m telescope by imaging with ~0.035 arcsec resolution. AOLI in the visible on the GTC 10.4m telescope has the diffraction limit of eight times better than HST of ~0.015 arcsec resolution, roughly 60 times the resolution when limited by atmospheric turbulence. There is every expectation that by making such high resolution images and spectra available more routinely many fields of astronomy will be revolutionised yet again.